\documentclass[12pt]{article}
\usepackage{epsfig,cite}
\newcount\timecount
\newcount\hours \newcount\minutes  \newcount\temp \newcount\pmhours
\hours = \time
\divide\hours by 60
\temp = \hours
\multiply\temp by 60
\minutes = \time
\advance\minutes by -\temp
\def\hour{\the\hours}
\def\minute{\ifnum\minutes<10 0\the\minutes
            \else\the\minutes\fi}
\def\clock{
\ifnum\hours=0 12:\minute\ AM
\else\ifnum\hours<12 \hour:\minute\ AM
      \else\ifnum\hours=12 12:\minute\ PM
            \else\ifnum\hours>12
                 \pmhours=\hours
                 \advance\pmhours by -12
                 \the\pmhours:\minute\ PM
                 \fi
            \fi
      \fi
\fi
}

\def\monthname{\relax\ifcase\month 0/\or January\or February\or
   March\or April\or May\or June\or July\or August\or September\or
   October\or November\or December\else\number\month/\fi}

\def\bold#1{\setbox0=\hbox{$#1$}%
     \kern-.025em\copy0\kern-\wd0
     \kern.05em\copy0\kern-\wd0
     \kern-.025em\raise.0433em\box0 }

\def\beq{\begin{equation}}
\def\eeq{\end{equation}}

\def\ga{\mathrel{\raise.3ex\hbox{$>$\kern-.75em\lower1ex\hbox{$\sim$}}}}
\def\la{\mathrel{\raise.3ex\hbox{$<$\kern-.75em\lower1ex\hbox{$\sim$}}}}
\def\gev{{\rm \, Ge\kern-0.125em V}}
\def\tev{{\rm \, Te\kern-0.125em V}}
\def\gyr{{\rm \, G\kern-0.125em yr}}

\def\gappeq{\mathrel{\rlap {\raise.5ex\hbox{$>$}}
{\lower.5ex\hbox{$\sim$}}}}
\def\lappeq{\mathrel{\rlap{\raise.5ex\hbox{$<$}}
{\lower.5ex\hbox{$\sim$}}}}
\def\Toprel#1\over#2{\mathrel{\mathop{#2}\limits^{#1}}}

\def\stau{\widetilde \tau}

\def\mpl{M_{\rm Pl}}
\def\mchi{m_{\tilde \chi}}

\def\m12{m_{1\!/2}}

\def\mstau{m_{\tilde{\ell}_1}}

\def\stau{\tilde{\tau}}
\def\mstau{m_{\tilde{\tau}}}
\def\mgrav{m_{3/2}}
\def\mpl{M_{P}}
\def\mchi{m_{\chi}}
\def\cosw{\cos \theta_W}
\def\sinw{\sin \theta_W}
\def\bea{\begin{eqnarray}}
\def\eea{\end{eqnarray}}

\begin{document}
\begin{titlepage}
\pagestyle{empty}
\baselineskip=21pt
\rightline{\tt hep-ph/0312262}
\rightline{CERN--TH/2003-310}
\rightline{UMN--TH--2225/03}
\rightline{FTPI--MINN--03/37}
\vskip 0.2in
\begin{center}
{\large {\bf Gravitino Dark Matter in the CMSSM}}
\end{center}
\begin{center}
\vskip 0.2in
{\bf John~Ellis}$^1$, {\bf Keith~A.~Olive}$^{2}$, {\bf Yudi~Santoso}$^{2}$ 
and {\bf Vassilis~C.~Spanos}$^{2}$
\vskip 0.1in
{\it
$^1${TH Division, CERN, Geneva, Switzerland}\\
$^2${William I. Fine Theoretical Physics Institute, \\
University of Minnesota, Minneapolis, MN 55455, USA}}\\
\vskip 0.2in
{\bf Abstract}
\end{center}
\baselineskip=18pt \noindent

We consider the possibility that the gravitino might be the lightest
supersymmetric particle (LSP) in the constrained minimal extension of the
Standard Model (CMSSM). In this case, the next-to-lightest supersymmetric
particle (NSP) would be unstable, with an abundance constrained by the
concordance between the observed light-element abundances and those
calculated on the basis of the baryon-to-entropy ratio determined using
CMB data. We modify and extend previous CMSSM relic neutralino
calculations to evaluate the NSP density, also in the case that the NSP is
the lighter stau, and show that the constraint  from late NSP decays is
respected only in a limited region of the CMSSM parameter space. In this
region, gravitinos might constitute the dark matter.

\vfill
\leftline{CERN--TH/2003-310}
\leftline{December 2003}
\end{titlepage}
\baselineskip=18pt

\section{Introduction}

If $R$ parity is conserved, the lightest supersymmetric particle (LSP) is
stable, and a possible candidate for the cold dark matter postulated by
astrophysicists and cosmologists~\cite{EHNOS}.  Most analyses of such
supersymmetric dark matter have assumed that the LSP is the partner of
some combination of Standard Model particles, such as the lightest
neutralino $\chi$, with an abundance calculated from the freeze-out of
annihilation processes in a thermal initial state. However, another
generic possibility is that the LSP is the gravitino ${\tilde
G}$~\cite{ekn} - \cite{feng}, whose relic abundance would get
contributions from the decays of the next-to-lightest supersymmetric
particle (NSP) and possibly other mechanisms.

As we discuss in more detail below, the lifetime of the NSP is typically
such that it decays between Big-Bang nucleosynthesis (BBN) and the `re-'
combination process when the cosmic microwave background (CMB) was
released from matter. Since NSP decays release entropy during this epoch, 
they are
constrained by the concordance of the observed light-element abundances
with BBN calculations assuming the baryon-to-entropy ratio inferred from
CMB observations. For a typical lifetime $\tau_{NSP} = 10^8$~s, the 
observed $^6$Li abundance implies ~\cite{CEFO}
\begin{equation}
{n_{NSP} \over n_\gamma} \; < \; 
5 \times 10^{-14} \left( { 100~{\rm GeV} \over m_{NSP} } \right)
\label{CEFO}
\end{equation}
before NSP decay, with the D/H ($^4$He) abundance providing a constraint
which is weaker by a factor of about 10 (20). 
Assuming a baryon-to-entropy ratio 
$\eta \equiv n_B / n_\gamma = 6.0  \times 10^{-10}$,  
in  agreement with the WMAP result 
$\eta = 6.1^{+0.3}_{-0.2} \times 10^{-10}$~\cite{wmap}, 
(\ref{CEFO}) implies the constraint $n_{NSP} / n_B 
< 10^{-4} ( 100~{\rm GeV} / m_{NSP})$ before the onset of NSP decay. To 
assess the power of this constraint, we re-express it in terms of 
$\Omega^0_{NSP} h^2$, the relic density that the NSP {\it would have} 
today, {\it if} it had not decayed:
\begin{equation}
\Omega^0_{NSP} h^2 \; < \; 10^{-2}  \; \Omega_B h^2 \simeq 2 \times 10^{-4},
\label{NSPoverB}
\end{equation}
where $\Omega_B h^2 \simeq 2 \times 10^{-2}$ is the present-day baryon 
density. However, the requirement (\ref{NSPoverB}) would be relaxed for a 
shorter-lived NSP~\cite{BBB}, as we discuss later.

In contrast, assuming that the lightest neutralino $\chi$ is the LSP, 
there have been
many calculations of $\Omega_\chi h^2$ in the constrained minimal
supersymmetric extension of the Standard Model (CMSSM), in which the
GUT-scale input gaugino masses $m_{1/2}$ and scalar masses $m_0$ are each
assumed to be universal~\cite{eos,cmssmnew,eoss,cmssmmap}. 
These calculations find generic strips of CMSSM
parameter space in which
\begin{equation}
\Omega_\chi h^2 \; \sim \; 5 \times \Omega_B h^2 \; \sim \; 0.1
\label{usual} 
\end{equation}
This is similar to the range of the cold dark matter density
$\Omega_{CDM} h^2$ favoured by astrophysicists and cosmologists, which is
one reason why neutralino dark matter has been quite popular. 

In this paper, we assume no {\it a priori} relation between $\mgrav$ and
the soft supersymmetry-breaking masses $m_{1/2}$ and $m_0$ of the
spartners of Standard Model particles in the CMSSM. This is possible in,
e.g., the framework of $N = 1$ supergravity with a non-minimal K\"ahler
potential~\cite{sugra}. In such a framework, the LSP might well be the
gravitino ${\tilde G}$. In this case, the NSP would likely be the lightest
supersymmetric partner of some combination of Standard Model particles,
such as the lightest neutralino $\chi$ or the lighter stau ${\tilde
\tau_1}$. Particularly in the $\chi$ NSP case, one might expect
$\Omega^0_{NSP} h^2$ to be near the range (\ref{usual}). Comparing this
with the condition (\ref{NSPoverB}) necessary for gravitino dark matter,
we see that, if $\tau_{NSP} = 10^8$~s, gravitino dark matter could be
possible only in rather different regions of the CMSSM parameter space,
where the NSP density is very suppressed compared with the usual $\chi$
density. Moreover, in this case, NSP decays alone could not provide enough
gravitinos, since they could only yield $\Omega_{3/2} h^2 <
\Omega^0_{NSP} h^2$, so there would need to be some supplementary
mechanism for producing gravitinos, if they were to provide all the cold
dark matter. For example, gravitino production during reheating after inflation 
could produce a sufficient abundance of gravitinos if the reheat temperature
is relatively large, $\sim {\cal O}(10^{10})$~GeV~\cite{BBB}.

The first step in our exploration of the gravitino dark
matter possibility is to calculate $\Omega^0_{NSP} h^2$ throughout the
$(m_{1/2}, m_0)$ planes for different choices of $\tan \beta$ and the sign
of $\mu$ in the CMSSM, assuming that the trilinear soft 
supersymmetry-breaking parameter $A_0 = 0$. In the regions where $m_\chi < 
m_{\tilde \tau_1}$,
this is essentially equivalent to the usual neutralino dark matter density
calculation. However, as we discuss below, this calculation must be
adapted in the region where $m_{\tilde \tau_1} < m_\chi$. Moreover, one
must take into account the possibility of a cosmological ${\tilde \tau_1}$
asymmetry, in which case the relic ${\tilde \tau_1}$ density would be
larger than that given by the standard freeze-out calculation. We next
compute the NSP lifetime and use the detailed constraints from
the abundances of the light elements as computed in  (\ref{CEFO}) for fixed
$\eta = 6 \times 10^{-10}$. This allows us to delineate the regions of the 
CMSSM $(m_{1/2}, m_0)$ planes where
gravitino dark matter appears possible. We find limited regions of the
$(m_{1/2}, m_0)$ planes that are allowed.  In these regions, the density
of relic gravitinos due to NSP decay is typically less than the range
favoured by astrophysics and cosmology. As noted above, supplementary 
mechanisms for gravitino production, such as thermal production in the early Universe,
might then enable gravitinos to constitute the cold dark matter.

\section{NSP Density Calculations}

In the framework of the CMSSM with a light gravitino discussed here, the
candidates for the NSP are the lightest partners of Standard Model particles.
In generic regions of CMSSM parameter space, these are the lightest neutralino
$\chi$ and the lighter stau ${\tilde \tau_1}$~\footnote{The lighter stop 
could also be the NSP if the trilinear coupling $A_0$ is 
large~\cite{stopco}, but here we fix $A_0 = 0$ for simplicity.}.
In regions where $\chi$ is the NSP, the calculation of the NSP density
$\Omega^0_{NSP} h^2$ is identical with that of $\Omega_{LSP} h^2$ in the CMSSM
with a heavier gravitino, and we can recycle standard results.

Extending these calculations of $\Omega^0_{NSP} h^2$ to regions where the
${\tilde \tau_1}$ is the NSP requires some modifications. Whereas the
Majorana $\chi$ is its own antiparticle, one must distinguish between the
${\tilde \tau_1}$ and its antiparticle ${\tilde \tau_1}^*$, and
calculate the sum of their relic densities. This requires a careful
accounting of the statistical factors in all relevant annihilation and
coannihilation processes. We have also made a careful treatment of the
regions where there is rapid ${\tilde \tau_1} - {\tilde
\tau_1}^*$ annihilation via Higgs poles, and a non-relativistic expansion
in powers of the NSP velocity is inadequate. Here our treatment follows
that of the neutralino LSP case in~\cite{ganis,eos,eoss}. 

It is important to note that one would, in general, expect a net 
${\tilde
\tau_1}$ asymmetry $\eta_{\tilde \tau_1} \equiv \lambda \, \eta_B$, where $\lambda \sim
{O}(1)$. This would be the expectation, for example, in
leptogenesis scenarios, and would also appear in other baryogenesis
scenarios, as a result of electroweak sphalerons. 
However, in the context of the MSSM, there exist ${\tilde \tau_1} {\tilde
\tau_1} \to \tau \tau$ annihilation processes which would bleed away any existing
lepton asymmetry stored in the ${\tilde \tau}$ sleptons, and the final relic 
density is given by the calculation described above.

\section{NSP Decays}

Using the standard $N = 1$ supergravity Lagrangian~\cite{cremmer,wess}, 
one can calculate the rates for the various decay channels of candidate 
NSPs to gravitinos.

The dominant decay of a $\chi$ NSP would be into a gravitino and a 
photon, for which we calculate the width
\beq
\Gamma_{\chi \to \tilde{G} \, \gamma}=
\frac{1}{16\pi} \, \frac{C^2_{\chi \gamma}}{\mpl^2} \, 
\frac{\mchi^5}{\mgrav^2}
\, \left( 1-\frac{\mgrav^2}{\mchi^2} \right)^3
\, \left( \frac{1}{3} + \frac{\mgrav^2}{\mchi^2} \right)
\eeq
where $C_{\chi \gamma}=(O_{1 \chi} \cosw + O_{2 \chi} \sinw)$
and $O$ is the neutralino diagonalization matrix,
$O^T \, {\cal M}_N \, O = {\cal M}_N^{diag}$. Note that
in this and the following equations $M_P \equiv 1/\sqrt{8 \pi G_N}$.

A $\chi$ NSP may also decay into a gravitino and a $Z$ boson, for which 
we calculate the rate
\bea
\Gamma_{\chi \to \tilde{G} \, Z} &=&
\frac{1}{16\pi} \, \frac{C^2_{\chi Z}}{\mpl^2} \, \frac{\mchi^5}{\mgrav^2}
\, {\cal F}(\mchi,\mgrav,M_Z) \nonumber \\
&\times& \left\{ \left( 1-\frac{\mgrav^2}{\mchi^2} \right)^2
\, \left( \frac{1}{3} + \frac{\mgrav^2}{\mchi^2} \right)
-\frac{M_Z^2}{\mchi^2} \, {\cal G}(\mchi,\mgrav,M_Z) \right\}
\eea
where $C_{\chi Z}=(-O_{1 \chi} \sinw + O_{2 \chi} \cosw$), and we use the
auxiliary functions
\beq
{\cal F}(\mchi,\mgrav,M_Z)=
\left[ \left(1-\left( \frac{\mgrav+M_Z}{\mchi} \right)^2 \right)
\,\left(1-\left( \frac{\mgrav-M_Z}{\mchi} \right)^2 \right) \right]^{1/2} 
\,,
\eeq
\beq
{\cal G}(\mchi,\mgrav,M_Z)= 1+
\frac{\mgrav^3}{\mchi^3}\, \left(-4+\frac{\mgrav}{3\, \mchi} \right)
 + \frac{M_Z^4}{3\, \mchi^4} - \frac{M_Z^2}{\mchi^2}
\left(1 - \frac{\mgrav^2}{3\, \mchi^2} \right) \,.
\eeq
Note that in the limit $M_Z \to 0$ we obtain
$\Gamma_{\chi \to \tilde{G} \, Z} \to \Gamma_{\chi \to \tilde{G} \gamma} $
by replacing $C_{\chi Z}$ with $C_{\chi \gamma}$.

Decays of a  $\chi$ NSP into a gravitino and a Higgs boson are also 
possible, with a rate
\bea
\Gamma_{\chi \to \tilde{G} \, h} &=&
\frac{1}{96\pi} \, \frac{C^2_{\chi h}}{\mpl^2} \, \frac{\mchi^5}{\mgrav^2}
\, {\cal F}(\mchi,\mgrav,m_h) \nonumber \\
&\times&  \left( \left( \frac{\mgrav}{\mchi} +1 \right)^2
     -\frac{m_h^2}{\mchi^2} \right)^2
 \left( \left( \frac{\mgrav}{\mchi} - 1 \right)^2
     -\frac{m_h^2}{\mchi^2} \right)
\label{decayh}
\eea
where $C_{\chi h}=(O_{4 \chi} \cos\alpha - O_{3 \chi} \sin\alpha)$.

Analogously, for the heavy Higgs boson $H$ we get
$\Gamma_{\chi \to \tilde{G} h} \to \Gamma_{\chi \to \tilde{G} H} $
by replacing $C_{\chi h}$ with
$C_{\chi H} \equiv (O_{4 \chi} \sin\alpha + O_{3 \chi} \cos\alpha)$,
and $m_h$ with $m_H$.
The corresponding formula for $\chi \to {\tilde G} + A$, where $A$ 
is the CP-odd Higgs boson in the MSSM, is also given by (\ref{decayh}), 
but with $m_h$ replaced by $m_A$, $\mgrav \to -\mgrav$ and
$C_{\chi h} \to C_{\chi A} \equiv (O_{4 \chi} \cos \beta + O_{3 \chi} \sin 
\beta)$.

Finally, the dominant decay of a ${\tilde \tau}$ NSP would be into a 
gravitino and a $\tau$, with the rate:
\beq
\Gamma_{\stau \to \tilde{G} \, \tau}=
\frac{1}{48\pi} \, \frac{1}{\mpl^2} \, \frac{\mstau^5}{\mgrav^2}
\, \left( 1- \frac{\mgrav^2}{\mstau^2} \right)^4 \,.
\eeq
where we have neglected the $O(m_\tau^2/m_{\tilde \tau_1}^2)$ terms.

\section{Effects of Gravitino Decay Products on Light-Element Abundances}

The effects of electromagnetic shower development between Big-Bang
Nucleosynthesis (BBN) and `re-'combination have been well studied, most
recently in~\cite{CEFO}, where the simplest case of $\chi \to {\tilde G} +
\gamma$ decays were considered. The late injection of electromagnetic
energy can wreak havoc on the abundances of the light elements.  
Energetic photons may destroy deuterium, destroy $^4$He (which may lead to
excess production of D/H), destroy $^7$Li, and/or overproduce $^6$Li. The
concordance between BBN calculations and the observed abundances of these
elements can be used to derive a limit on the density of any decaying
particle. In general, this limit will depend on both the baryon asymmetry
$\eta_B$, which controls the BBN predictions, and on the life-time of the
decaying particle $\tau_X$. For a fixed value $\eta_B = 6 \times
10^{-10}$, as suggested by CMB observations, the bounds derived from
Fig.~8(a)  in~\cite{CEFO} may be parameterized approximately as
\begin{equation}
y < 0.13 \, x^2 -2.85 \, x +3.16 \, ,
\label{CEFOtau}
\end{equation}
where $y \equiv \log(\zeta_X/{\rm GeV}) \equiv \log(m_X n_X/n_\gamma/{\rm GeV})$ and $x \equiv
\log(\tau_X/{\rm s})$, for the electromagnetic decays of particles $X$ with
lifetimes $10^{12}~{\rm s} > \tau_{NSP} \ga 10^{4}$~s. In our subsequent
analysis we use the actual data corresponding to the limit in~\cite{CEFO}
in order to delineate the allowed regions of the $(m_{1/2}, m_0)$ planes,
but (\ref{CEFOtau}) may help the reader understand qualitatively our
results.

The other NSP decay modes listed above inject electrons, muons and hadrons
into the primordial medium, as well as photons. Electromagnetic showers
develop similarly, whether they are initiated by electrons or photons, so
we can apply the analysis of~\cite{CEFO} directly also to electrons.
Bottom, charm and $\tau$ particles decay before they interact with the
cosmological medium, so new issues are raised only by the interactions of
muons, pions and strange particles. In fact, if the NSP lifetime exceeds
about $10^4$~s, these also decay before interacting, and the problem
reduces to the purely electromagnetic case. In the case of a shorter-lived
NSP, we would need to consider also hadronic interactions with the
cosmological medium~\cite{kohri}, which would strengthen the limits on
gravitino dark matter that we derive below on the basis of electromagnetic
showers alone. In the following, we do not consider regions of the
$(m_{1/2}, m_0)$ planes where $\tau_{NSP} < 10^{4}$~s.

It is sufficient for our purposes to treat the decays of $\mu, \pi$ and
$K$ as if their energies were equipartitioned among their decay products.  
In this approximation, we estimate that the fractions of particle energies
appearing in electromagnetic showers are $\pi^0 : 100$\%, $\mu: 1/3$,
$\pi^\pm: 1/4$, $K^\pm:  0.3$, $K^0: 0.5$. Using the measured decay
branching ratios of the $\tau$, we then estimate that $\sim 0.3$ of its
energy also appears in electromagnetic showers. In the case of generic
hadronic showers from $Z$ or Higgs decay, we estimate that $\sim 0.6$ of
the energy is electromagnetic, due mainly to $\pi^0$ and $\pi^\pm$
production.

Our procedure is then as follows. First, on the basis of a freeze-out
calculation, we calculate the NSP relic density $\Omega_{X}^0 h^2
= 3.9 \times 10^7 \ {\rm GeV}^{-1} \  \zeta_X$. Next, we use the calculated life-time $\tau_X$
to compute the ratio of the relic density to the limiting value,
$\zeta_X^{CEFO}$ provided by the analysis of~\cite{CEFO}, taking into
account the electromagnetic energy decay fractions estimated above.  
Finally, we require
\beq
r \equiv {\zeta_X \over \zeta_X^{CEFO}} < 1.
\label{r}
\eeq

\section{Results}

As compared to the case of CMSSM dark matter usually discussed, in the
case of gravitino dark matter one must treat $\mgrav$ as an additional
free parameter, unrelated {\it a priori} to $m_0$ and $m_{1/2}$. We 
incorporate the LEP constraint on $m_h$ in the same
way as in~\cite{eoss}~\footnote{For simplicity, we do not show the LEP
constraints on $m_{\chi^\pm}$ and $m_{\tilde e}$, which do not impinge on
the regions of parameters allowed by other constraints.}, and it appears 
as a nearly vertical
(red) dot-dashed line in each of the following figures. Regions excluded by
measurements of $b \to s \gamma$ are shaded dark (green). For
reference, the figures also
display the strips of the $(m_{1/2}, m_0)$ planes where $0.094 < 
\Omega_{NSP}^0 < 0.129$. This density is the same as $\Omega_{LSP} h^2$ in 
a standard CMSSM analysis with a heavy gravitino, extended to include the
unphysical case where the ${\tilde \tau_1}$ is the LSP. We note the
familiar `bulk' regions and coannihilation `tails', as well as
rapid-annihilation `funnels' for large $\tan \beta$ \cite{funnel,ganis}. 
If these figures were extended to larger $m_0$, 
there would also be `focus-point' regions~\cite{hbrsb,focus}.

We now summarize our principal results, describing the interplay of these
constraints with those associated specifically with gravitino dark matter,
studying the $(m_{1/2}, m_0)$ planes for three choices of $\tan \beta$ and
the sign of $\mu$: (1) $\tan \beta = 10$, $\mu > 0$, (2) $\tan \beta = 35$,
$\mu < 0$, and (3) $\tan \beta = 50$, $\mu > 0$. 
In each case, we consider four
possibilities for
$\mgrav$: two fixed values 10~GeV and 100~GeV, and two fixed ratios
relative to $m_0$: $\mgrav = 0.2 \, m_0$ and $m_0$ itself. If $\mgrav \gg
m_0$, the ${\tilde G}$ is typically not the LSP, and this role is played
by the lightest neutralino $\chi$, as assumed in most analyses of the
CMSSM. In each $(m_{1/2}, m_0)$ plane, we display as a (purple) dashed line
the limit where the density of relic gravitinos from NSP decay becomes
equal to the highest cold dark matter density allowed by WMAP and other
data at the 2-$\sigma$ level, namely $\Omega_{3/2} h^2 < 0.129$: only
regions below and to the right of this contour are allowed in our
analysis.

\begin{figure}
\begin{center}
\mbox{\epsfig{file=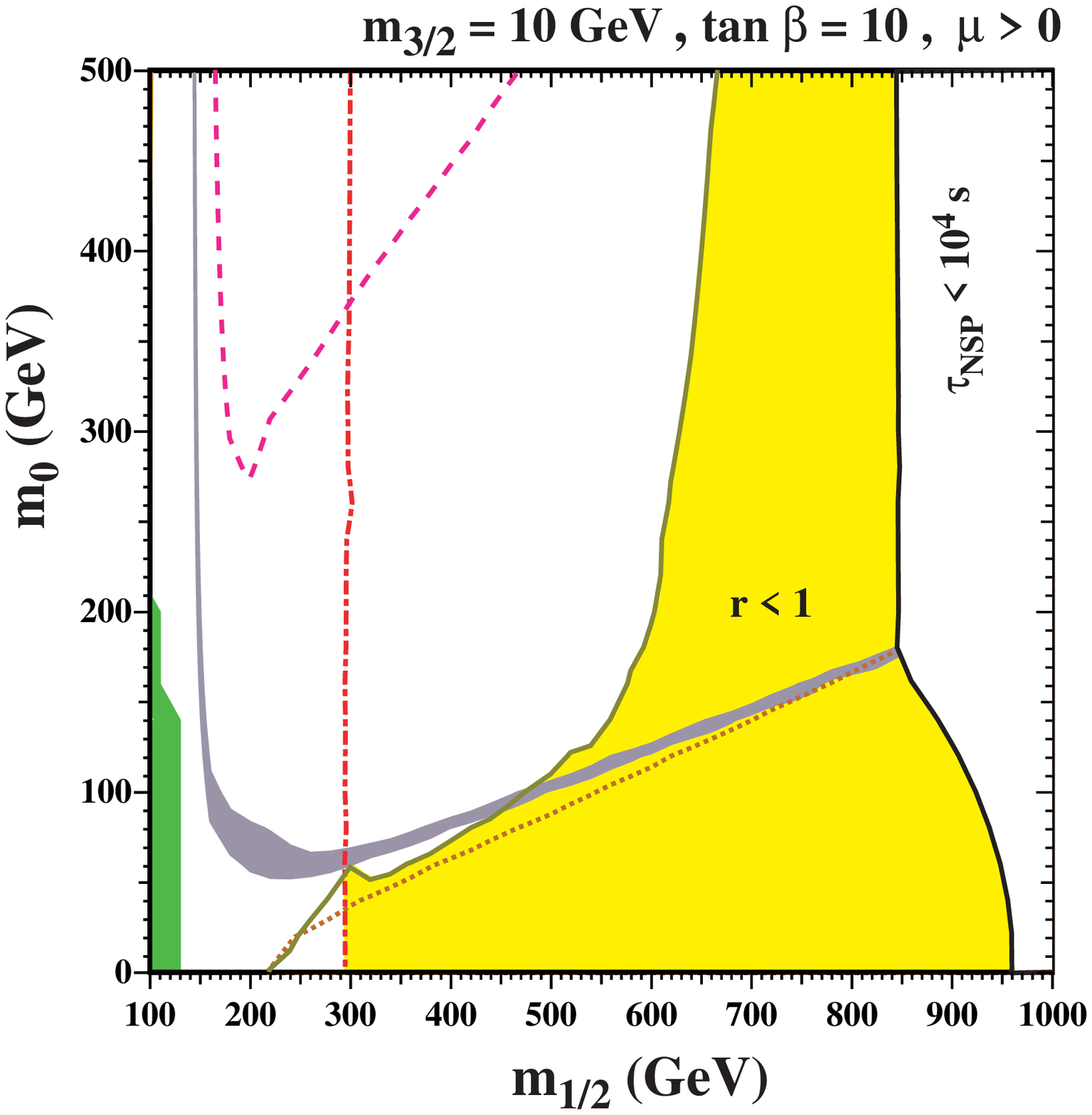,height=7cm}}
\mbox{\epsfig{file=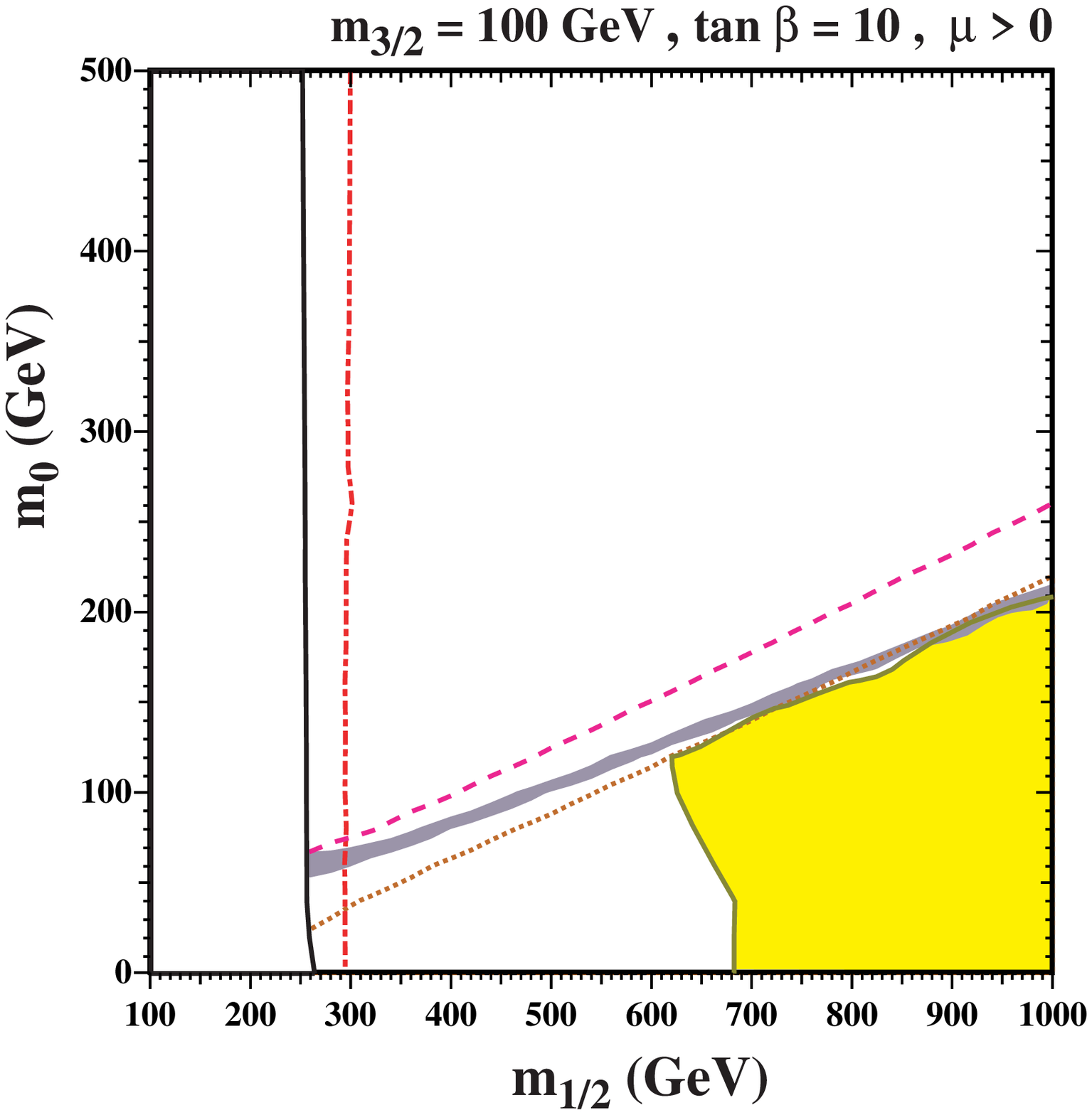,height=7cm}}
\end{center} 
\begin{center}
\mbox{\epsfig{file=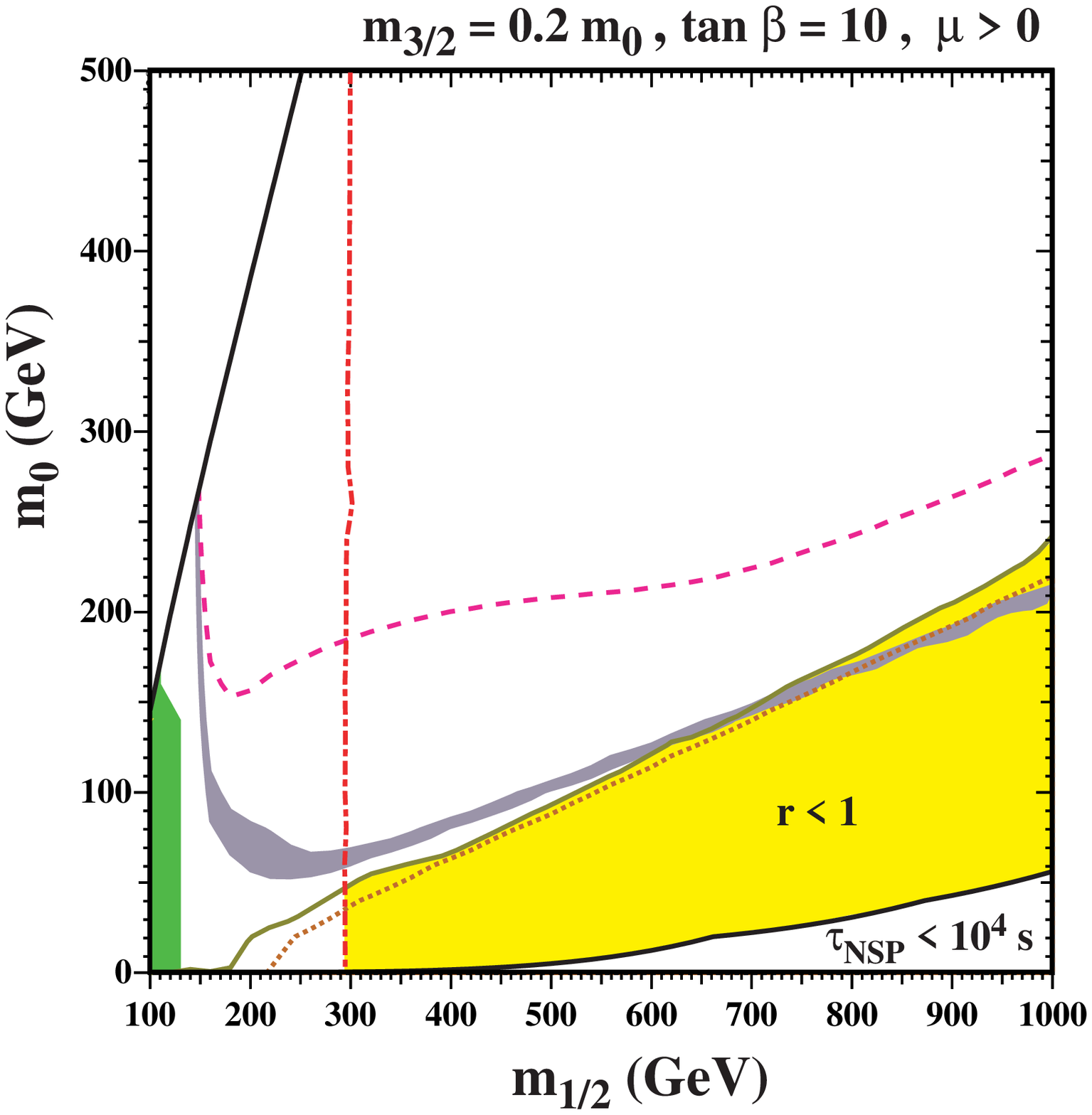,height=7cm}}
\mbox{\epsfig{file=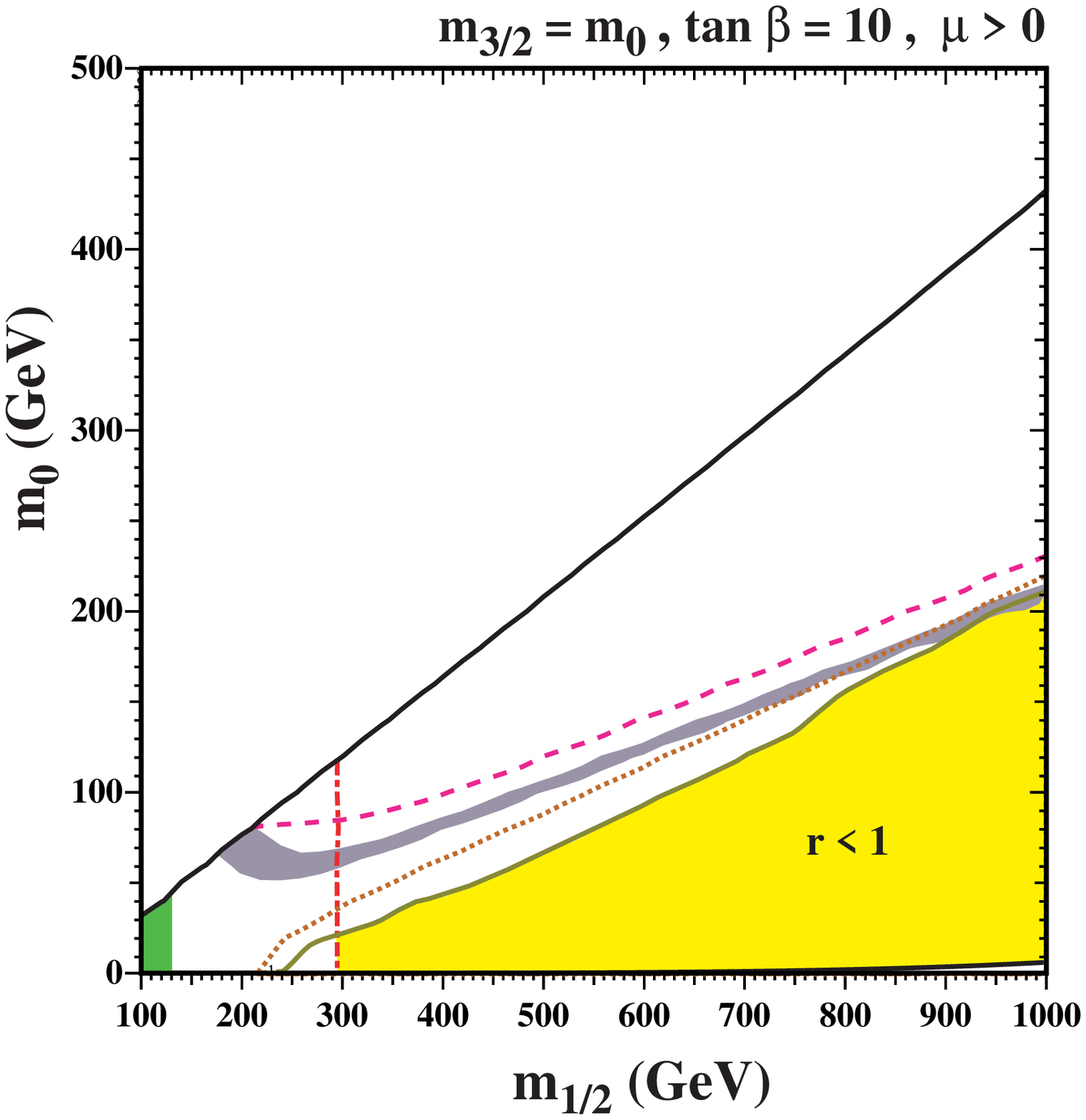,height=7cm}}
\end{center} 
\caption{\it
The $(m_{1/2}, m_0)$ planes for $\tan \beta =10, \mu > 0$ and the choices
(a) $\mgrav = 10$~GeV, (b) $\mgrav = 100$~GeV, (c) $\mgrav = 0.2 m_0$ and
(d) $\mgrav = m_0$. In each panel, we show $m_h = 114$~GeV 
calculated using {\tt FeynHiggs}~\cite{FeynHiggs}, as a near-vertical 
(red)  dot-dashed line, the region excluded by $b \to s \gamma$ is darkly shaded
(green), and the region where the NSP density before decay lies in the 
range $0.094 < \Omega^0_{NSP} h^2 < 0.129$ is medium shaded (grey-blue).
The (purple) dashed line is the contour where gravitinos 
produced in NSP decay have $\Omega_{3/2} h^2 = 0.129$, and the grey
(khaki) solid line ($r=1$) is the constraint on NSP decays provided by Big-Bang 
nucleosynthesis and CMB observations. The light (yellow) shaded region 
is allowed by all the constraints.  The contour where $m_\chi = 
m_{\tilde \tau_1}$ is shown as a (red) diagonal dotted line. Panels (a) and (c)
show as a black solid line the contour beyond which $\tau_{NSP} < 
10^4$~s, the case not considered here. Panels (b), (c), and (d)  show black 
lines to whose left the gravitino is no longer the LSP.} 
\label{fig:NSP10p} 
\end{figure}

Fig.~\ref{fig:NSP10p} displays the $(m_{1/2}, m_0)$ planes for $\tan \beta
= 10$ and $\mu > 0$~\footnote{The case $\tan \beta = 10$ and $\mu < 0$ is
very similar, with the exception that the $b \to s \gamma$ constraint is
more important.}. Panel (a) displays the choice $\mgrav = 10$~GeV, in
which case the LSP is the ${\tilde G}$ throughout the displayed region of
the $(m_{1/2}, m_0)$ plane. Above and to the left of the  (purple)  
dashed line, the relic density $\Omega_{3/2} h^2$ of gravitinos
yielded by NSP decay exceeds the 2-$\sigma$ upper limit on the cold dark
matter density, 0.129, imposed by WMAP and other cosmological data. This
region is therefore excluded. In the regions below the (purple)  
dashed line, the relic ${\tilde G}$ density might be increased so as to
provide the required cold dark matter density if there were significant
thermal gravitino production, in addition to that yielded by NSP decay.

The light-element constraint on NSP decays is shown as the grey
(khaki) solid line corresponding to $r = 1$, where $r$ is defined in
(\ref{r}). Regions to the right and below this line are allowed by this
constraint.  Here, and in the remaining figures below, the
region which satisfies the abundance constraint is labelled $r < 1$. There
is a solid (black) line with $m_{1/2} \sim 800$~GeV which indicates where
$\tau_{NSP} = 10^{4}$~s. To the right of this line, $\tau_{NSP} <
10^{4}$~s, the case we do not consider here because additional constraints
due to hadronic decays must be included, so this region is left
blank~\footnote{This line would disappear to larger $m_{1/2}$ already for
$\mgrav = 20$~GeV.}. Here and in subsequent figures, the region that is 
allowed by all the constraints is shaded in light (yellow) color.  

We see that there is an extended strip between the grey (khaki) solid line
and the solid (black) line. This strip is truncated 
above $m_0
\simeq 650$ GeV, because the relic density of gravitinos from NSP decay
becomes too large. This is true up to $\sim 2900$ GeV, where the relic
density drops as we approach the focus-point region. Here a small allowed
region opens up as the $r = 1$ curve bends towards lower values of
$m_{1/2}$.
The allowed strip broadens in the low-$m_0$ region where
$m_{\tilde \tau_1} < m_\chi$, below the dotted (red) line where $m_\chi =
m_{\tilde \tau_1}$.  In this region, gravitino dark matter is permitted.

Turning now to panel (b) of Fig.~\ref{fig:NSP10p}, where the choice
$\mgrav = 100$~GeV is made, we see a near-vertical black line at $m_{1/2}
\sim 250$~GeV: the gravitino is the LSP only to its right. The $\tau_{NSP}
= 10^{4}$~s line has disappeared to larger $m_{1/2}$, and is not shown. In
this case the $\Omega_{3/2} h^2$ constraint is much more important than in
panel (a), forcing $m_0$ to be relatively small, simply because $\mgrav$
is larger. The only region allowed by the light-element constraint on NSP
decays is in the bottom right-hand corner, in the region where the
${\tilde \tau_1}$ is the NSP. 

In panel (c) of Fig.~\ref{fig:NSP10p}, for $\mgrav = 0.2 m_0$, there is
also a black line to whose right the ${\tilde G}$ is the LSP, which is now
diagonal, and the $\Omega_{3/2} h^2$ constraint is similar to that in
panel (b). Most of the region allowed by the light-element constraint on
NSP decays is in the region where the ${\tilde \tau_1}$ is the NSP, though
a sliver of parameter space runs above the dotted curve.  

Finally, in panel (d) of Fig.~\ref{fig:NSP10p}, where now $\mgrav = m_0$,
the ${\tilde G}$ constraint is more powerful, as is the $\Omega_{3/2} h^2$
constraint, and the region finally allowed by the light-element constraint
on NSP decays is again in the ${\tilde \tau_1}$ region. 

\begin{figure}
\begin{center}
\mbox{\epsfig{file=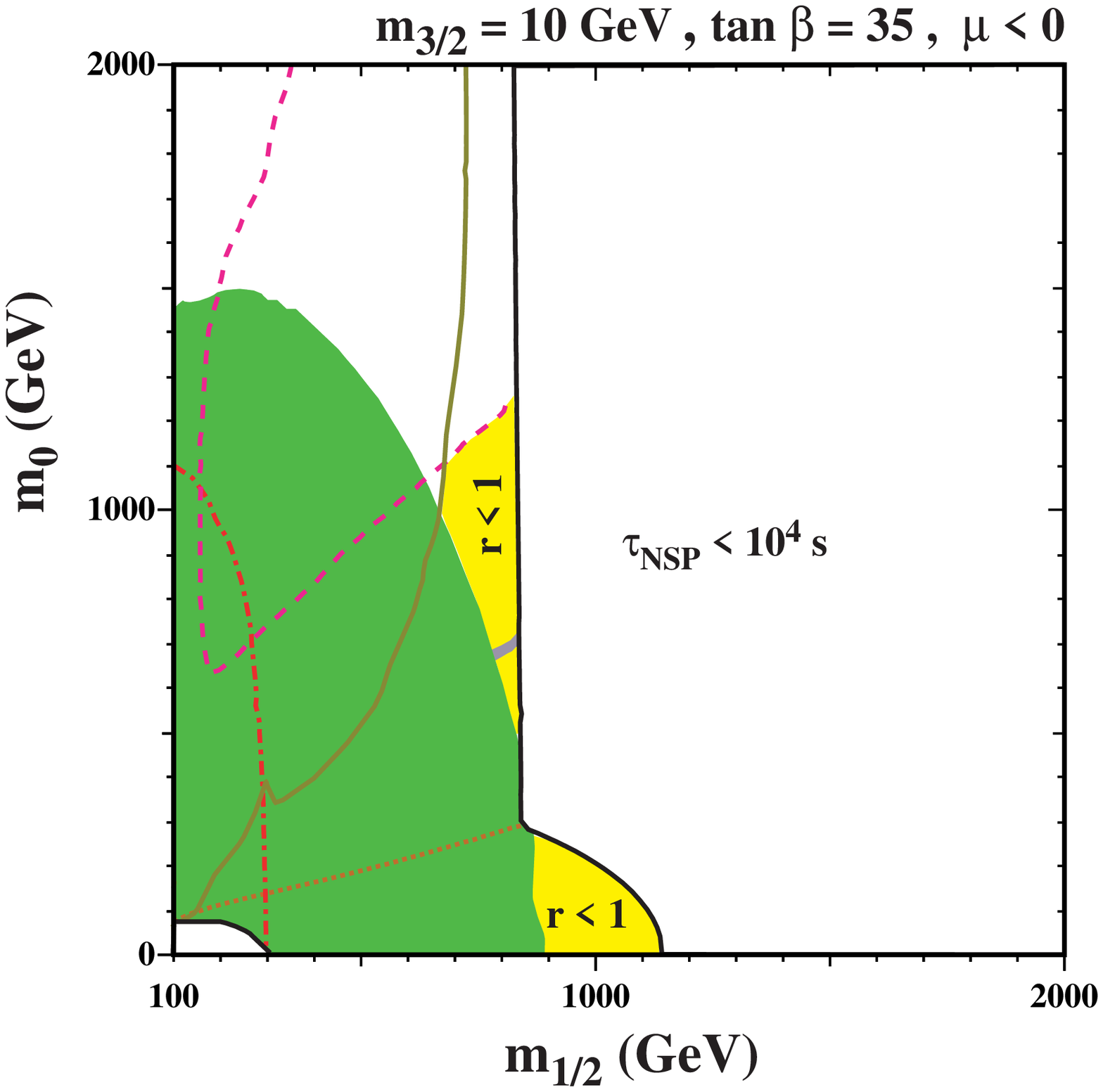,height=7cm}}
\mbox{\epsfig{file=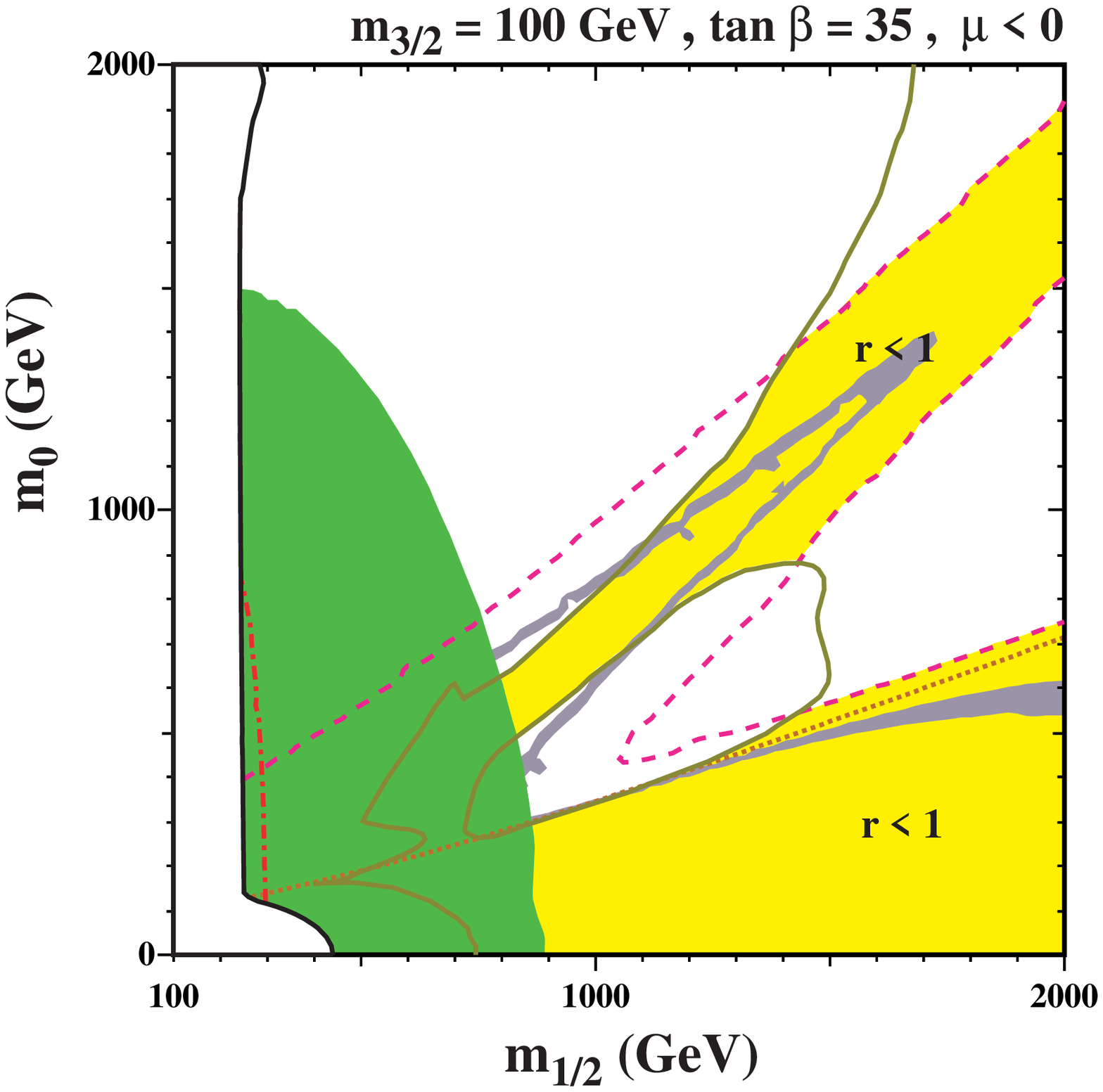,height=7cm}}
\end{center} 
\begin{center}
\mbox{\epsfig{file=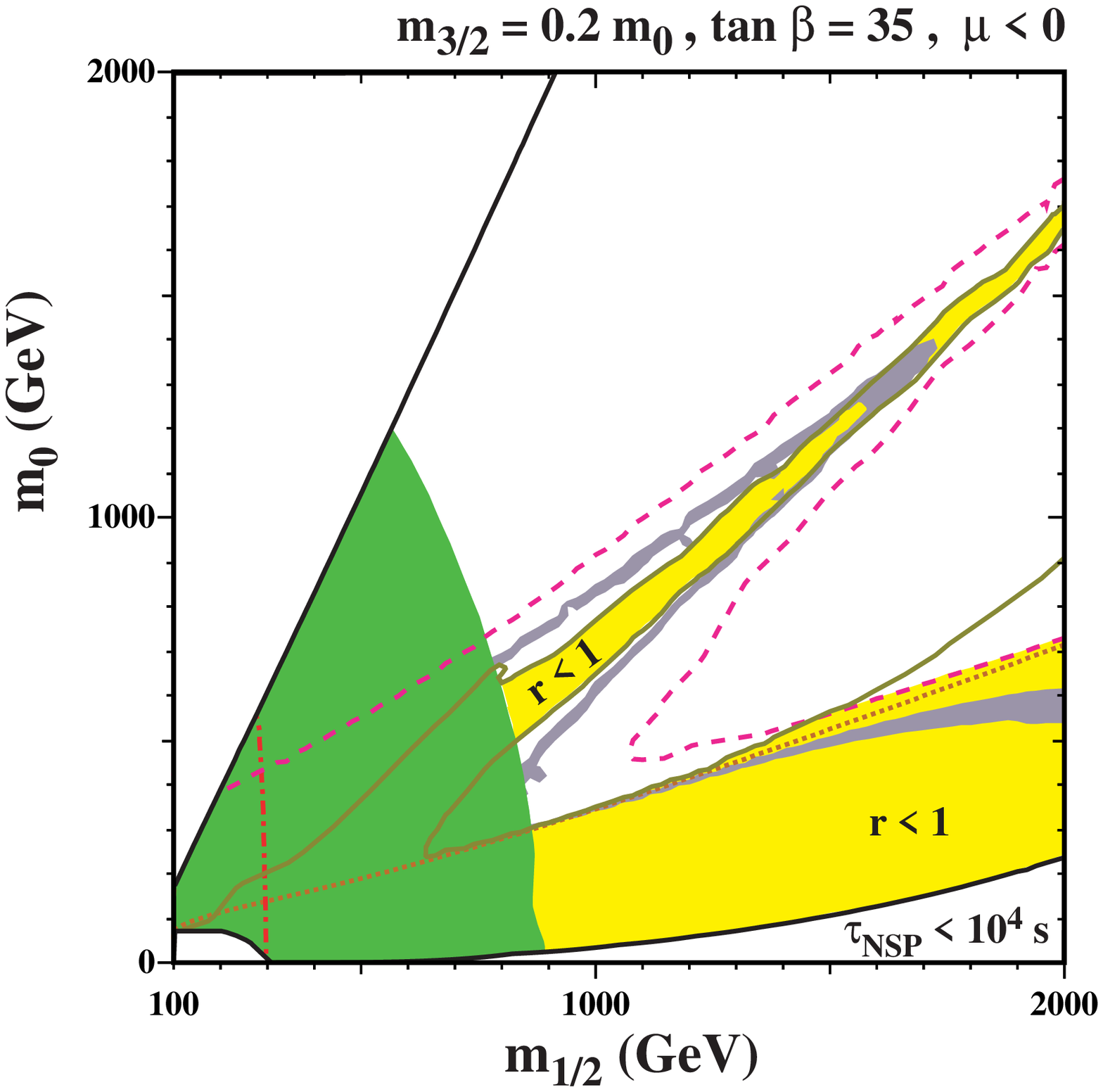,height=7cm}}
\mbox{\epsfig{file=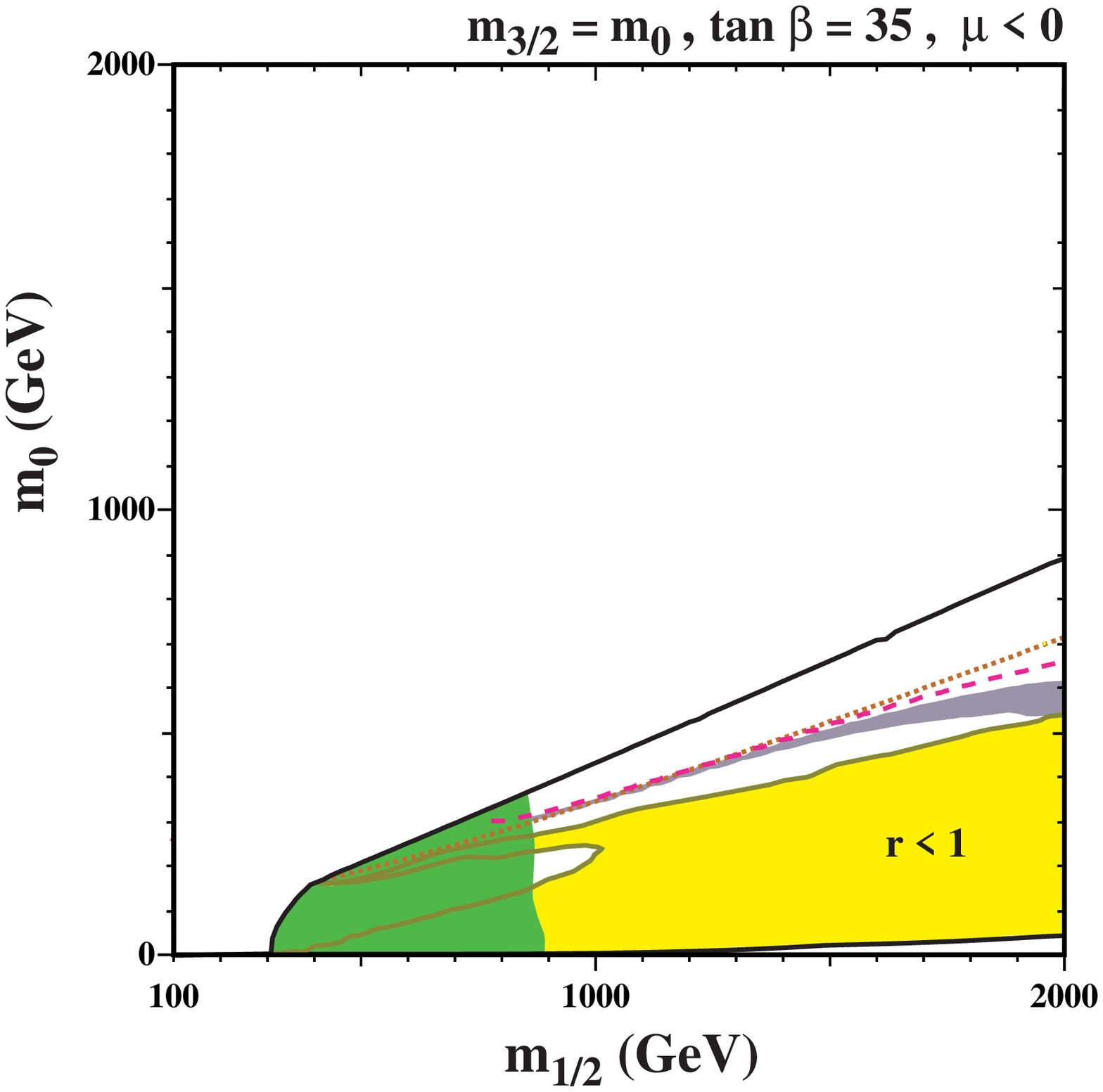,height=7cm}}
\end{center} 
\caption{\it
As in Fig.~\ref{fig:NSP10p}, for
$\tan \beta = 35$ and $\mu < 0$  and the choices
(a) $\mgrav = 10$~GeV, (b) $\mgrav = 100$~GeV, (c) $\mgrav = 0.2 m_0$ and
(d) $\mgrav = m_0$. The light (yellow) shaded regions are allowed by all 
the constraints.
}
\label{fig:NSP35n}
\end{figure}

Fig.~\ref{fig:NSP35n} displays a similar array of $(m_{1/2}, m_0)$ planes
for the case $\tan \beta = 35$ and $\mu < 0$. In the case where $m_{\tilde
\tau_1} = 10$~GeV, shown in panel (a), the most significant change
compared with panel (a) of Fig.~\ref{fig:NSP10p} is that the $b \to s
\gamma$ constraint is more important, whilst the $\Omega_{3/2} h^2$, NSP
decay and $\tau_{NSP}$ constraints do not change so much. The net result
is to leave disconnected parts of both the $\chi$ and ${\tilde \tau_1}$
regions that are allowed by all the constraints.

The most obvious new feature in panel (b) of Fig.~\ref{fig:NSP35n} is the
rapid-annihilation funnel, which affects both the $\Omega_{3/2} h^2$ and
NSP decay constraints. The former acquires a strip extending to large
$m_{1/2}$ and $m_0$, whereas the latter would have allowed a region at
large $m_{1/2} \gappeq 1500$~GeV that is excluded by $\Omega_{3/2} h^2$.
Combining this and the NSP decay constraint, we again find two
disconnected allowed regions, one in the $\chi$ NSP region and one that is
almost entirely in the ${\tilde \tau_1}$ NSP region. 

The rapid-annihilation funnel is also very apparent in panel (c) of
Fig.~\ref{fig:NSP35n}, which displays the case $m_{\tilde \tau_1} = 0.2
m_0$, where again a strip allowed by both the $\Omega_{3/2} h^2$ and NSP
decay constraints extends to large $m_{1/2}$ and $m_0$. There are again
disconnected allowed regions in the $\chi$ and (mainly) the ${\tilde
\tau_1}$ NSP region. Note that this is constrained at large $m_{1/2}$ and
small $m_0$ by the $\tau_{NSP}$ constraint.
Finally, in panel (d) of Fig.~\ref{fig:NSP35n}, for $m_{\tilde \tau_1} =
m_0$, the region allowed by the ${\tilde G}$ LSP, $\Omega_{3/2} h^2$
and NSP decay constraints is restricted to the part of the $(m_{1/2},
m_0)$ plane where the ${\tilde \tau_1}$ is the NSP. 

\begin{figure}
\begin{center}
\mbox{\epsfig{file=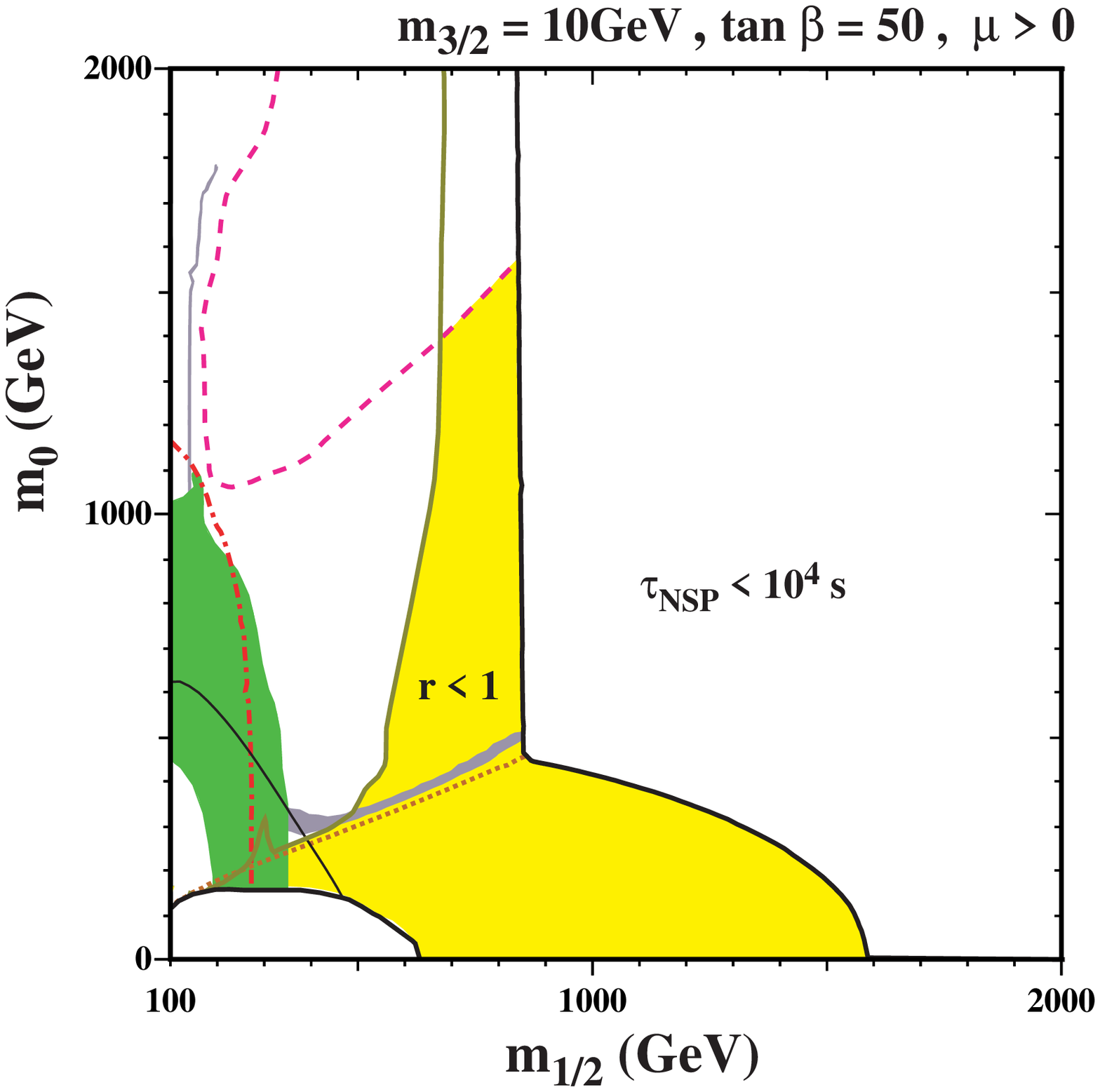,height=7cm}}
\mbox{\epsfig{file=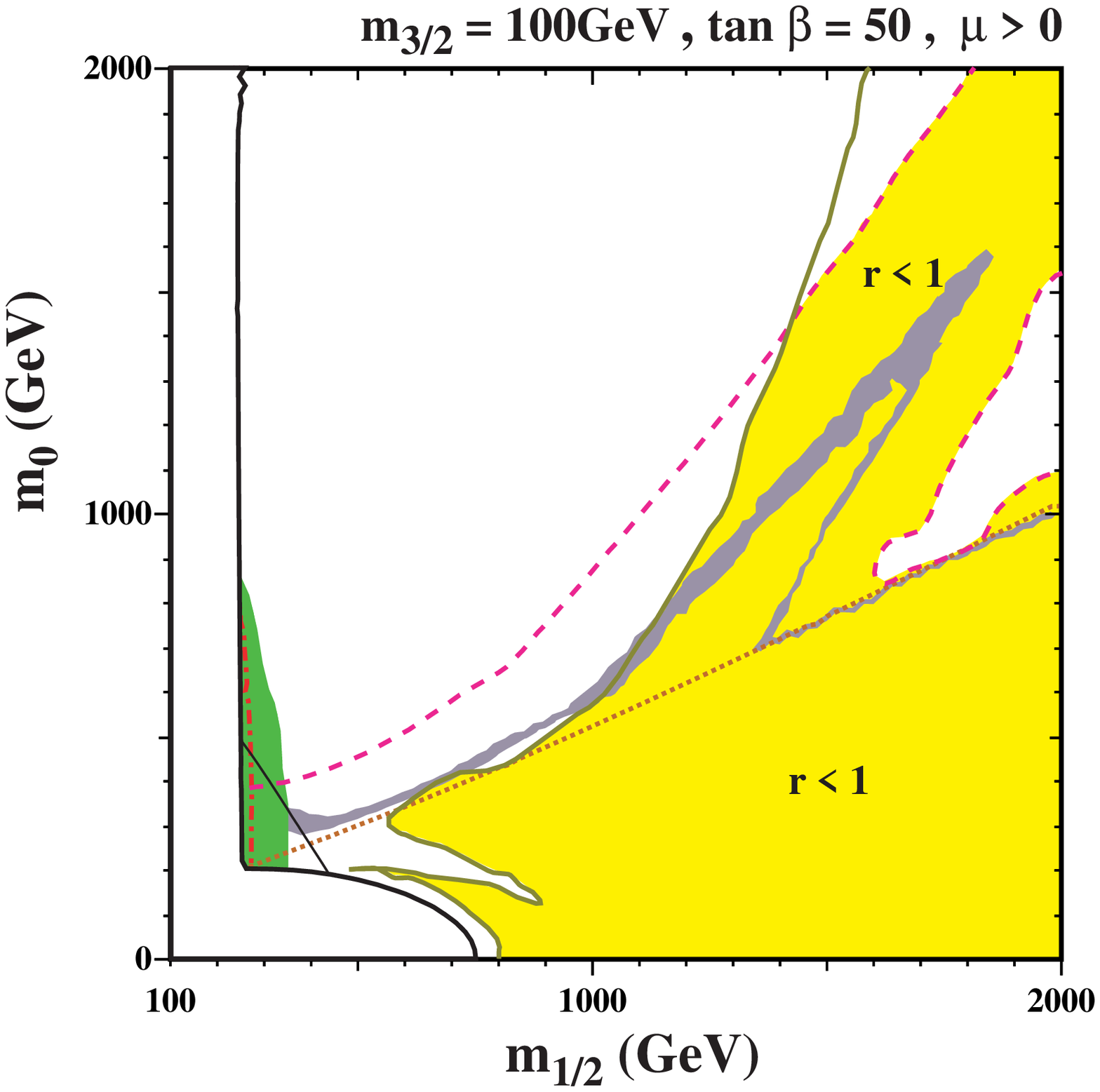,height=7cm}}
\end{center} 
\begin{center}
\mbox{\epsfig{file=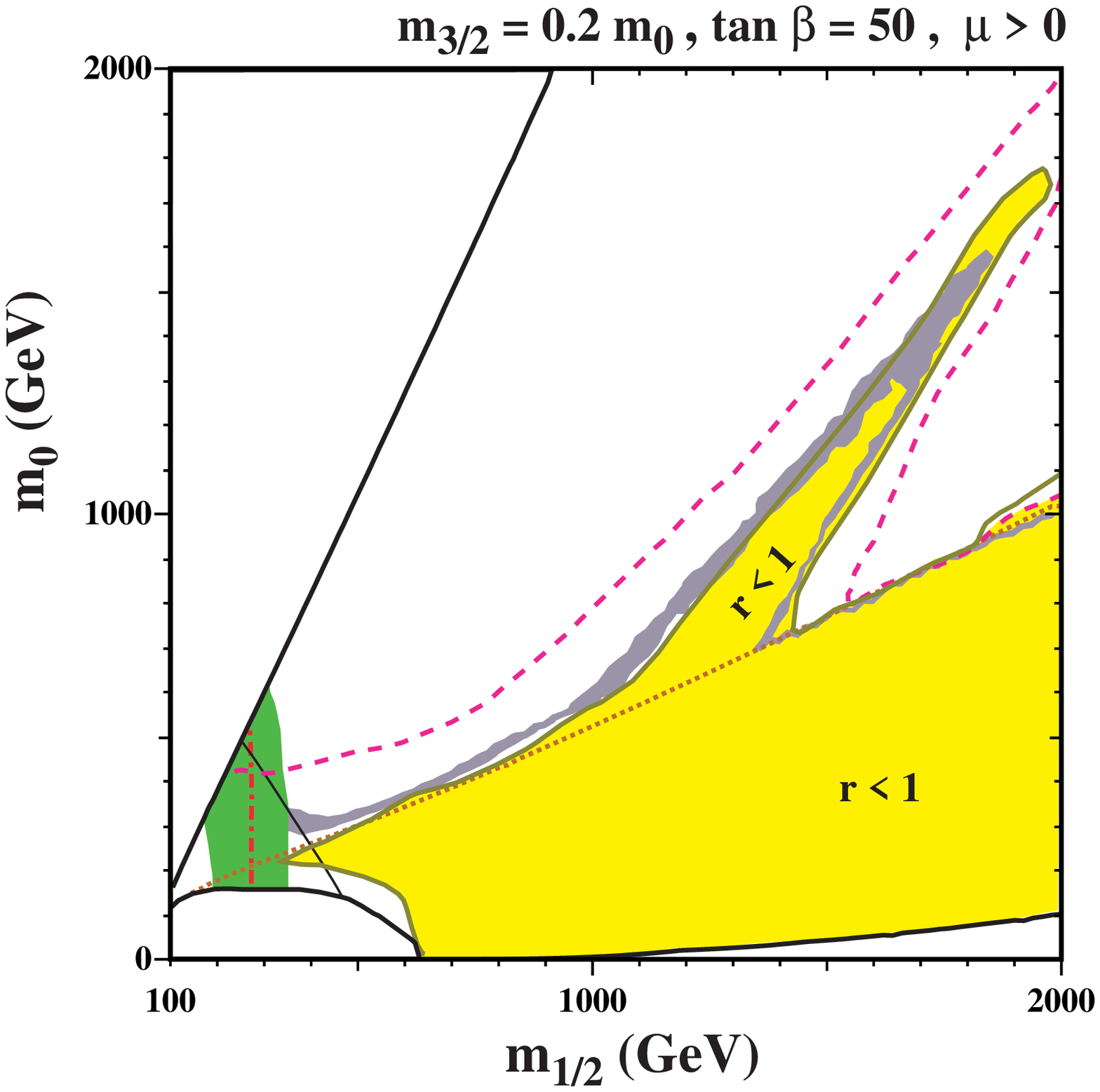,height=7cm}}
\mbox{\epsfig{file=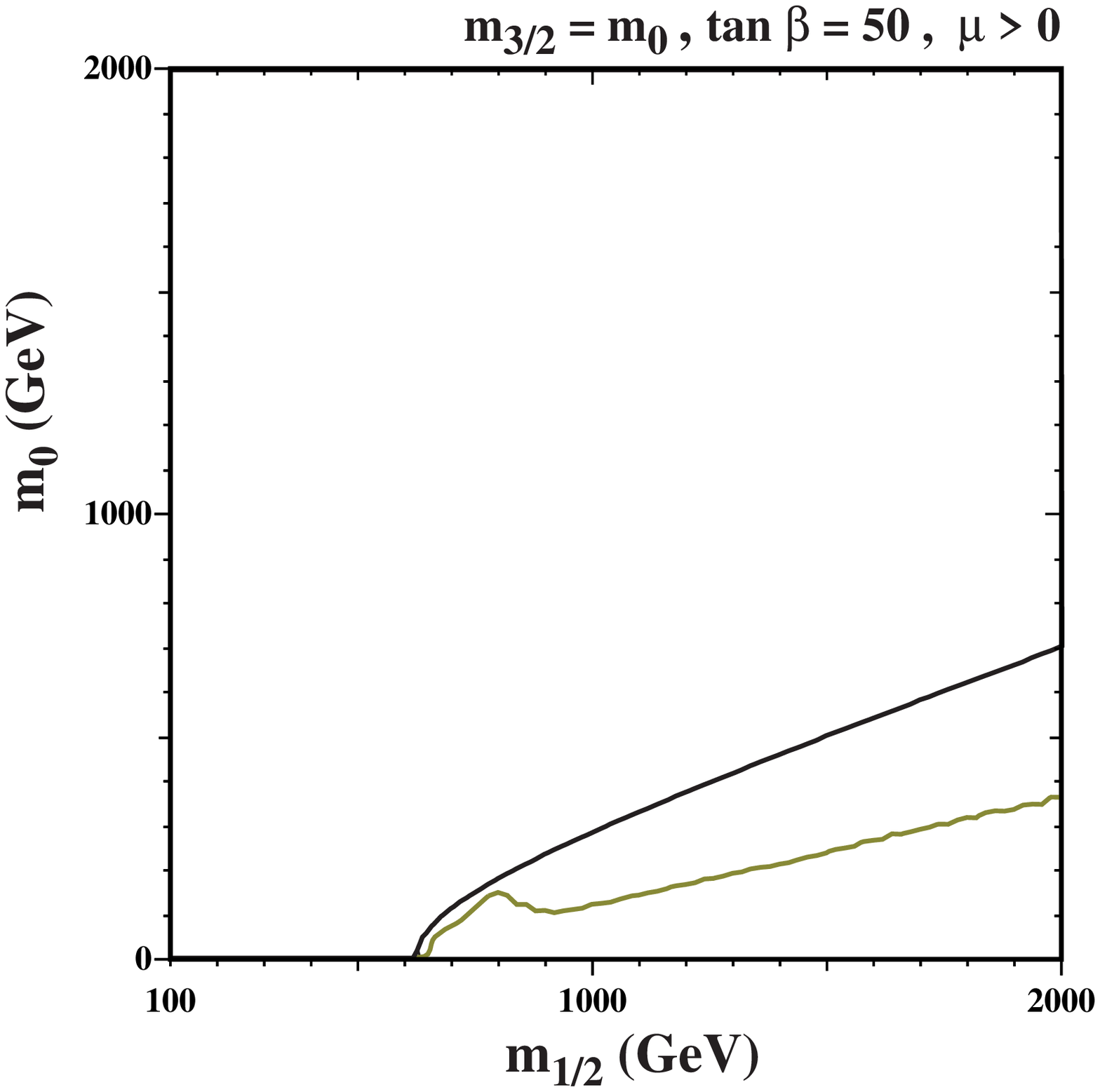,height=7cm}}
\end{center} 
\caption{\it
As in Figs.~\ref{fig:NSP10p} and \ref{fig:NSP35n}, for
$\tan \beta = 50$ and $\mu > 0$  and the choices
(a) $\mgrav = 10$~GeV, (b) $\mgrav = 100$~GeV, (c) $\mgrav = 0.2 m_0$ and
(d) $\mgrav = m_0$. In addition to the quantities plotted in the earlier 
figures, here we also plot grey solid lines 
where $a_\mu = 44.5 \times 10^{-10}$, 
which cut off at small $m_0$ the allowed regions in panels (a) and (c).
The light (yellow) shaded regions are allowed by all the constraints.
}
\label{fig:NSP50p}
\end{figure}

Fig.~\ref{fig:NSP50p} displays a similar array of $(m_{1/2}, m_0)$ planes
for the case $\tan \beta = 50$ and $\mu > 0$. The general features of the
planes have some similarities to those for $\tan \beta = 35$ and $\mu <
0$. There are differences in the interplays between the $\Omega_{3/2} h^2$
and NSP decay constraints, but an important difference is the relative
weakness of the $b \to s \gamma$ constraint. This has the consequence that
allowed $\chi$ and ${\tilde \tau_1}$ regions are connected for $\tan \beta
= 50$ and $\mu > 0$. It is interesting to note that this is the only case
where the putative constraint imposed by the muon anomalous magnetic
moment $a_\mu$ impinges on the allowed region, as shown in panels (a) and
(c). 

We have seen in the above examples that many of the allowed parts of the
$(m_{1/2}, m_0)$ planes are confined to regions where the NSP is the
${\tilde \tau_1}$.

\section{Conclusions}

We have analyzed in this paper the possibility of gravitino cold dark
matter within the CMSSM framework. Combining accelerator and cosmological
constraints, particularly those from $b \to s \gamma$, $\Omega_{3/2}
h^2$ and the light-element constraint on NSP decays, we have found allowed
regions in the $(m_{1/2}, m_0)$ planes for representative values of $\tan
\beta$ and the sign of $\mu$ and different values of $\mgrav$. Standard
calculations of the NSP density before decay based on freeze-out from
equilibrium yield allowed regions where either the lightest neutralino
$\chi$ or the lighter stau ${\tilde \tau_1}$ may be the NSP. 

One limitation of our analysis is that it is restricted to $\tau_{NSP} >
10^{4}$~s, in order to avoid issues related to the hadronic interactions
of NSP decay products before they decay. Also, in this paper we have not
discussed at much length what part of parameter space may be allowed in
the focus-point region. 
Finally, we have analyzed here only a few examples
of the possible relationship between $\mgrav$ and the CMSSM parameters
$m_0$ and $m_{1/2}$.

For these and other reasons, there are still many important issues to 
analyze concerning the possibility of gravitino dark matter. We have shown 
in this paper that such a possibility certainly exists, and that the 
allowed domains of parameter space are not very exceptional. We consider 
that gravitino dark matter deserves more attention than it has often 
received in the past. In particular, this possibility should be borne in 
mind when considering the prospects for collider experiments, since the 
allowed regions of the $(m_{1/2}, m_0)$ are typically rather different 
from those normally analyzed in the CMSSM. {\it Vive la diff{\'e}rence!}

\vspace*{1cm}
\noindent{ {\bf Acknowledgments} } \\
\noindent 
We would like to thank Rich Cyburt for providing us with BBN analysis data, and
Stefan Groot Nibbelink for helpful conversations. Y.S. would like to thank
Richard Arnowitt for useful discussions. 
We would also like to thank Koichi Hamaguchi
for constructive comments regarding the original version of the manuscript.
We thank Shufang Su and Fumihiro Takayama 
for pointing out errors in our paper, and Takeo Moroi for
clarifying the formula 4.31 in his thesis, hep-ph/9503210.  
The work of K.A.O., Y.S., and V.C.S. was supported in part
by DOE grant DE--FG02--94ER--40823.

\end{document}